\documentclass[preprintnumbers,showpacs,amsmath,amssymb,floatfix,9pt,prd,twocolumn,
superscriptaddress,nofootinbib]{revtex4}
\usepackage{latexsym}
\usepackage{epsfig}
\usepackage{amssymb}

\begin{document}

\title{Exact interior solutions in (2 + 1) dimensional  spacetime.}

\author{ Farook Rahaman}
\email{  rahaman@iucaa.ernet.in} \affiliation{Department of
Mathematics, Jadavpur University, Kolkata 700 032, West Bengal,
India}

\author{Piyali Bhar}
\email{piyalibhar90@gmail.com } \affiliation{Department of
Mathematics, Jadavpur University, Kolkata 700 032, West Bengal,
India}

\author{Ritabrata Biswas}
\email{biswas.ritabrata@gmail.com  } \affiliation{Indian
Institute of Engineering Sceince and Technology Shibpur
(Formerly, Bengal Engineering and Science University Shibpur),
Howrah 711 013, West Bengal, India}

\author{A. A. Usmani}
\email{anisul@iucaa.ernet.in} \affiliation{Department of Physics,
Aligarh Muslim University, Aligarh 202 002, Uttar Pradesh, India}

\begin{abstract}
\textbf{Abstract:}   We provide a new class of exact solutions for
the interior in (2 + 1) dimensional spacetime. The solutions
obtained for the perfect fluid model both with and without
cosmological constant ($\Lambda$) are found to be regular and
singularity free.  It assume very simple analytical forms that  help
us to study the various physical properties of the configuration.
Solutions without $\Lambda$ are found to be physically acceptable.
  \end{abstract}

\pacs{04.50.-h, 04.50.Kd, 04.20.Jb}

\maketitle

\section{Introduction}
Study of exact solutions of Einstein's Field Equations is an
important part of the theory of General Relativity. This importance
is not only from more formal mathematical aspects associated with
the theory (e.g.  the classification of space-times) but  also from
the  growing importance  of the  application of general relativity
to astrophysical phenomena.  For example, exact solutions may offer
physical insights that numerical solutions cannot.  Present time
trend of analyzing different aspects of black hole (BH) solutions
did lead us to grow our interests in cleaner $(2+1)$ dimensional
gravity. Discovery of BTZ BH~\cite{Bandos1} ignited the light first.
Through this $(2+1)$ dimensional model if we need to explore the
foundations of classical and quantum gravity we would not find any
Newtonian limit and no propagating degrees of freedom will arise. In
literature, very easy-to-find works in this aspect comprise the
study of quasi normal modes of charged dilaton BHs in $(2+1)$
dimensional solutions in low energy string theory with asymptotic
anti de-Sitter space times~\cite{Fernanado1}. Hawking radiation from
covariant anomalies in $(2+1)$ dimensional BHs~\cite{Nam1} is
another beautiful example. Lastly, we must also name the study of
branes with naked singularities analogous to linear or planar
defects in crystals and showing that zero branes in AdS space times
are ``negative mass BHs!"~\cite{Zanelli1}. Taking charged gravastars
as an alternative to charged BHs in $(2+1)$ AdS space times is
already investigated~\cite{Rahman1}. Extensions of BTZ BH solutions
with charge are also available in the literature. These are obtained
by employing nonlinear Born Infield electrodynamics to eliminate the
inner singularity~\cite{Mazharimousavi1}. The non-static charged BTZ
like BHs in $(N+1)$ dimensions have also  been studied \cite{Ghosh1}
which in its static limit, for $N=2$, reduces to $(2+1)$ BTZ BH
solutions.

Study of interior solutions in $(2+1)$ dimension~\cite{Rahman2}
shows that even the noncommutative-geometry-inspired BTZ BH is not
free from any singularity. Study of interior solutions are farely
found in literature. For example, solutions of C. Wolf~\cite{wolf}
and S. Yazadjiev~\cite{yaz}, solutions in te framework of
Brans-Dicke theory of gravity by S.M.Kozyrev \cite{kozyrev} and  new
class of solutions corresponding to BTZ exterior spacetime by Sharma
{\it et. al.}~\cite{Sharma}, which is regular at the centre and it
satisfies all the physical requirements except at the boundary where
the authors propose a thin ring of matter content with negative
energy density so as to prevent collapsing. The discontinuity of the
affine connections at the boundary surface provide the above matter
confined to the ring. Such a stress-energy tensor is not ruled out
from the consideration of Casimir effect for massless fields.

The purpose of the present work is to find exact interior solutions for
perfect fluid model both with and without cosmological constant, $\Lambda$.
The motivation for doing so is provided by the
fact that the assumption of equation of state (EoS), $p=m\rho$, which
seems to be very reasonable for describing the matter distribution
in the study of relativistic   objects like stars~\cite{ayan,cruz},
wormholes~\cite{fr1,kim} and gravastars~\cite{Rahman1,fr3}.

The structure of our work is as follows: In sec~(\ref{Ein_Field}),
we derive  required Einstein equations. Sec.~ \ref{Interior})
constitutes of different interior solutions for various cases of EoS.
Lastly, in sec.~(\ref{conclusion}) a brief conclusion is provided.

\section{Einstein field equations in  $(2 + 1)$ dimension}\label{Ein_Field}
 We  take the static metric to describe the interior region  of a $(2+1)$
dimensional space time as
 \begin{equation}\label{metric}
 ds^{2} = -e^{2\nu(r)}dt^{2}+ e^{2\mu(r)}dr^{2} + r^{2} d\theta^{2},
 \end{equation}
where $\nu(r)$ and $\mu(r)$ are the two unknown metric functions. We
take the perfect fluid   form of the energy momentum tensor
\begin{equation}\label{stress}
T_{ij}=  diag(-\rho,p,p),
\end{equation}
where $\rho$ is energy density and  $p$ is pressure.
Einstein's field equations with a
cosmological constant, $\Lambda$, for the space-time metric
(Eq.(\ref{metric})) together with the energy momentum tensor given in
Eq.(\ref{stress}) may be written as
\begin{eqnarray}\label{field_eq_1}
 2\pi\rho +\Lambda&=& \mu^{\prime}e^{-2\mu(r)}/r,\\
 2\pi p -\Lambda&=& \nu^{\prime}e^{-2\mu(r)}/r,\\
 2\pi p -\Lambda&=& e^{-2\mu}\left({\nu^{\prime\prime}}+{\nu^{\prime 2}}-\nu^{\prime}\mu^{\prime}\right).
 \end{eqnarray}
Here superscript `$\prime$' denotes the derivative with respect to
$r$. Assuming $G=c=1$, the generalized Tolman-Oppenheimer-Volkov
(TOV) equation may be written as
\begin{equation}\label{continuity}
\left(\rho+p\right)\nu^{\prime }+{p^{\prime }}   =0,
 \end{equation}
which represents conservation equations in $(2+1)$ dimensions.

We take the EoS of the form
\begin{equation}\label{EoS_1}
p  =m \rho,
 \end{equation}
where $m$ is EoS parameter.

\section{Interior solutions }\label{Interior}
We first obtain interior solutions without any cosmological
constant, thereby taking $\Lambda$=0.  Latter on, we generalize our
study to non-zero value of $\Lambda$. We choose various cases of EoS
parameter for both the choices of $\Lambda$.

\subsection{With no cosmological constant ($\Lambda=0$)}
\subsubsection{$~0~< m~< ~1~$}
For $\Lambda= 0$, the field equations (3)-(6) become
 \begin{eqnarray}\label{8}
  2 \pi \rho&=&\mu'e^{-2\mu}/r,\\
  2 \pi m\rho&=&\nu'e^{-2\mu}/r,\\
  2 \pi m\rho&=& e^{-2\mu}(\nu'^{2}+\nu''-\mu'\nu').
 \end{eqnarray}
The TOV equation (11)  takes the form,
\begin{equation}\label{EoS2}
(\rho+m \rho)\nu'+m \rho'=0,
\end{equation}
Equation (\ref{EoS2}) yields
\begin{equation}\label{12}
\rho^{m}e^{(1+m)\nu}=C.
\end{equation}
Solving equations (8) and (9),  we get
\begin{equation}\label{13}
 \nu=m\mu +A~~.
 \end{equation}
 Equating Eq.(9) with Eq.(10),  we get
\begin{equation}\label{14}
e^{\mu}=e^{\nu}\nu'/r.
\end{equation}
Now, solving equations (\ref{13}) and (\ref{14}),  we obtain
\begin{eqnarray}
\nu&=&
\frac{A}{1-m}-\frac{m}{1-m}ln\left\{\frac{1-m}{m}\left(B-\frac{r^{2}}{2}
\right)  \right\},\\
\mu&=&
\frac{1}{1-m}\left[A-ln\left\{\frac{1-m}{m}\left(B-\frac{r^{2}}{2}
\right)  \right\}  \right],\\
\rho&=& \frac{1}{2 \pi m} e^{-\frac{2A}{1-m}}\left[\frac{1-m}{m}\left(
B-\frac{r^{2}}{2} \right)\right]^{\frac{1+m}{1-m}}.
\end{eqnarray}
Here $C$, $A$ and $B$ are integration constants.

For the consistency of solutions, the constants should follow the
constraint equation,
\begin{equation}
A=m \ln(2 \pi m)+\ln C~~.
\end{equation}
These solutions  are regular at the center. The central density
is given by
\begin{equation}
\rho_c= 1/(2 \pi m) e^{-\frac{2A}{1-m}}\left[B(1-m)/m
\right]^{\frac{1+m}{1-m}}.
\end{equation}
The interior solution is valid up to the radius $r<\sqrt{2B}$.
For a physically meaningful solution the radial and tangential
pressure should be decreasing function of r.   From  equation (17),
we find
  \begin{equation}
\frac{d\rho}{dr}=\frac{1}{m} \frac{dp}{dr}
 < 0  ,
\end{equation}
which gives density and pressure as decreasing functions of r.
 At $r=0$, one can get
 \begin{equation}
\frac{dp}{dr}=0,~~ \frac{d\rho}{dr}=0 ~~{\rm and}~~
\frac{d^2\rho}{dr^2}=\left[\frac{(1-m)B}{m}\right]
^{\frac{2m}{1-m}} <0
\end{equation}
which support maximality of central density and radial central
pressure. Here, density and pressure decrease radially outward as
shown in FIG.~1.\\
\\

\begin{figure}[htbp]
    \centering
        \includegraphics[scale=.24]{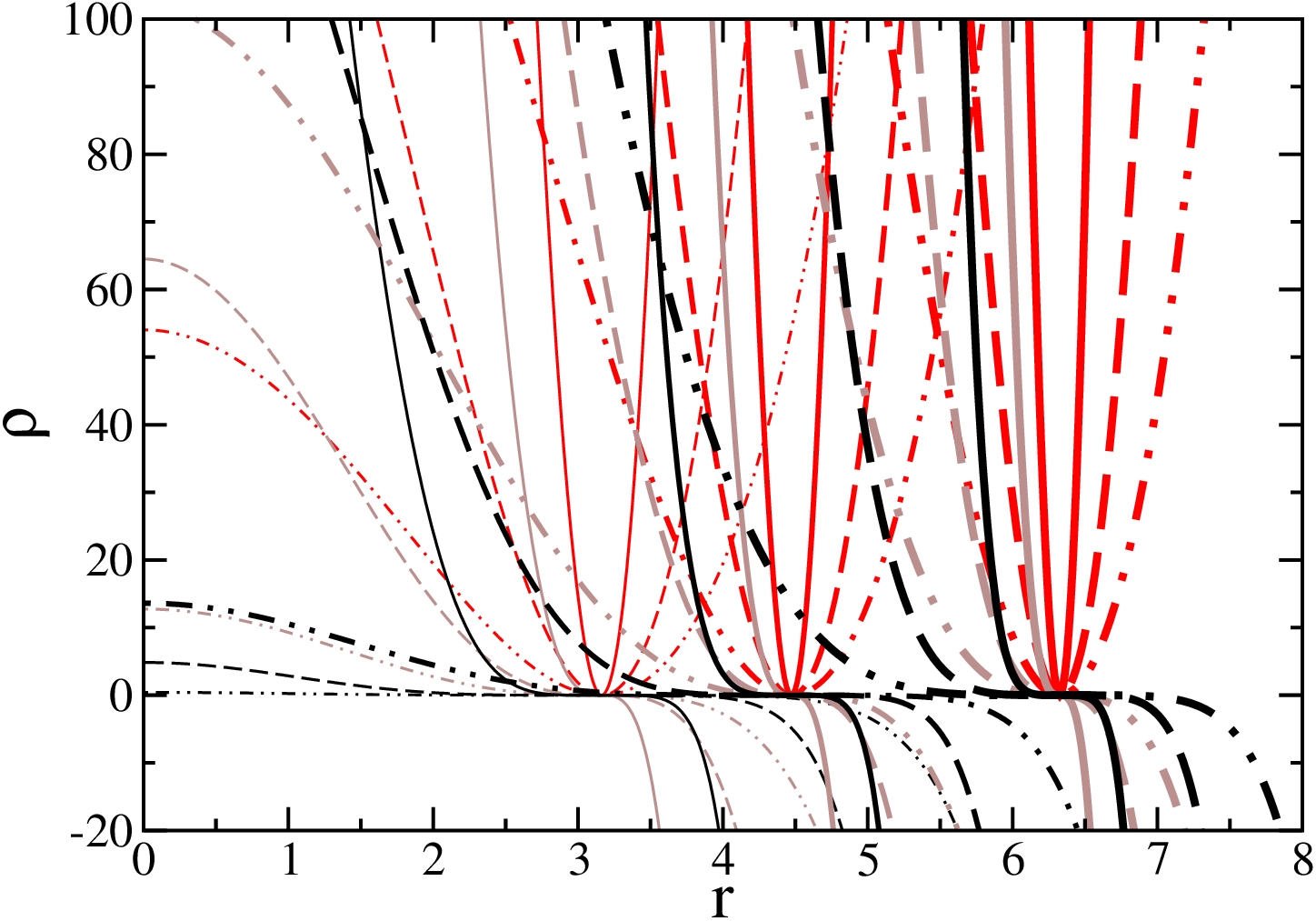}
       \caption{  Variation of the energy-density($\rho$)    in the
  interior region. Description of curves are as follows:
The red, brown and black colors represent $m$=1/3, $m$=1/2 and
$m$=2/3, respectively.  For all these, solid, dashed and chain lines
represent $C=0.25$, 0.5 and 0.75, respectively. Thin, thick and
thickest lines correspond to B=5, 10, and 20, respectively. }
    \label{fig:1}
\end{figure}
The above TOV (Eq.~11) may be re-written as
\begin{eqnarray}
 \frac{M_G\left(\rho+p\right)}{r}e^{\frac{\mu-\nu}{2}}+\frac{dp}{dr}
=0,
\end{eqnarray}
where $M_G=M_G(r)$ is the gravitational mass inside a sphere of
radius $r$ and is given by Tolman-Whittaker formula, which may be derived
from  field equations,
\begin{equation}
M_G(r)=re^{\frac{\nu-\mu}{2}}\nu^{\prime}\label{eq26}.
\end{equation}

This modified form of  TOV equation indicates  the equilibrium
condition for the fluid sphere  subject to the gravitational and
hydrostatic  forces,
\begin{equation}
 F_g+ F_h =0,
\end{equation}
where
\begin{small}
\begin{eqnarray}
F_g &=&  \nu^{\prime}\left(\rho+p\right)
=r\frac{1+m}{2\pi m}e^{-\frac{2A}{1-m}}\left[\frac{1-m}{m}\left(B-\frac{r^{2}}{2}\right)\right]^{\frac{2m}{1-m}}, \\
F_h &=&  \frac{dp}{dr}=-F_g.
\end{eqnarray}
\end{small}
  The profiles of $F_g$ and  $F_h$   for the specific values of the parameters
are shown in FIG.~2 which provides the
information  about the static equilibrium   due to the combined
effect of   gravitational and hydrostatic forces.

Mass, $M(r)$, within a radius $r$, is calculated as
\begin{small}
\begin{eqnarray}
M(r) &=& \int_0^r 2\pi \rho \tilde{r}d\tilde{r} =
\frac{1}{2}e^{-\frac{2A}{1-m}}\left[B\left(\frac{1-m}{m}\right)\right]^{\frac{2}{1-m}} \nonumber \\
&-&\frac{1}{2}e^{-\frac{2A}{1-m}}\left[\left(B-\frac{r^{2}}{2}\right)\left(\frac{1-m}{m}\right)\right]^{\frac{2}{1-m}}.
\end{eqnarray}
\end{small}
\\
\\
\begin{figure}[htbp]
    \centering
        \includegraphics[scale=.24]{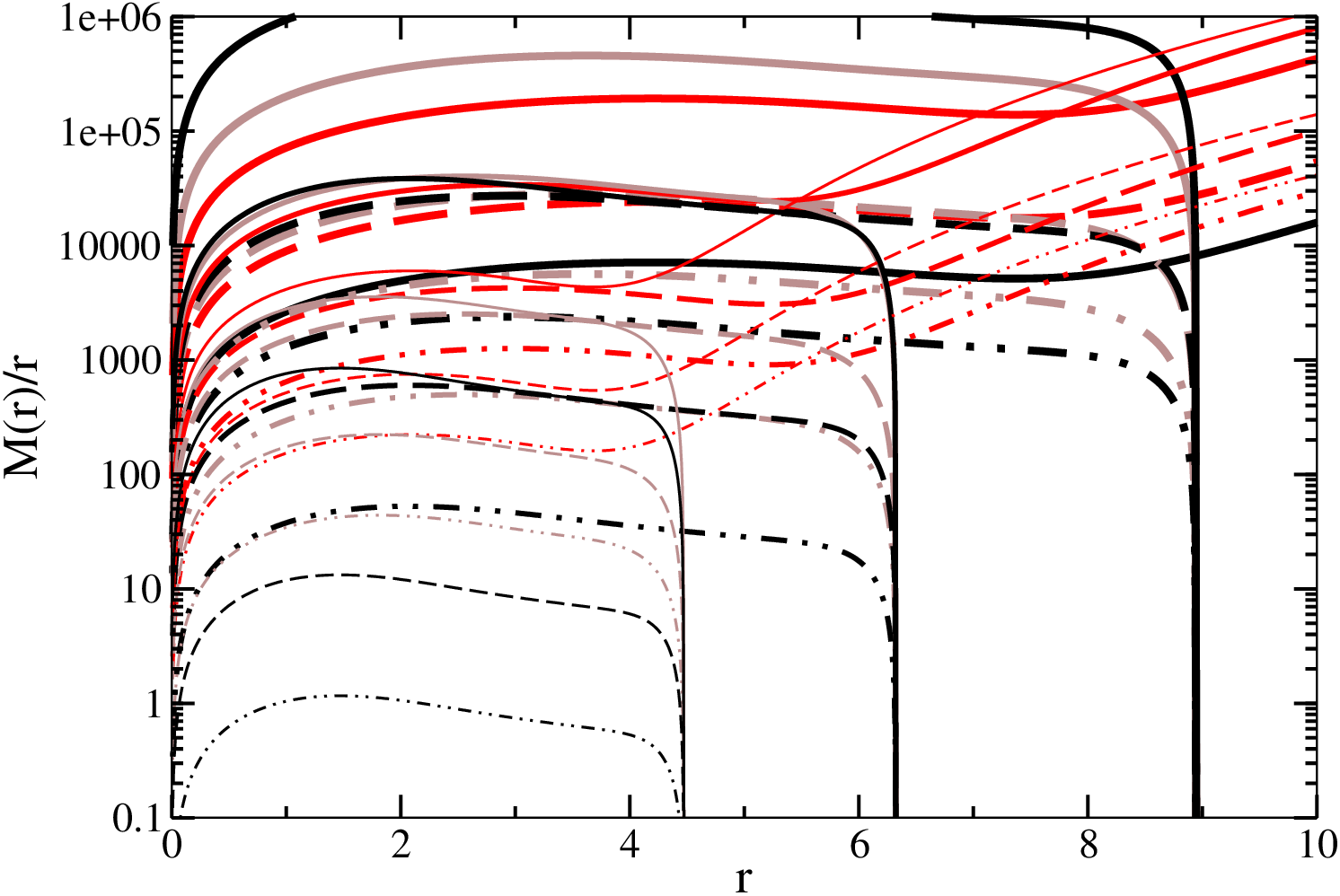}
       \caption{    Variation of compactness( $u=\frac{M(r)}{r}$  )  in the
  interior region. Description of curves is the same as in FIG.~1. }
    \label{fig:2}
\end{figure}

\begin{figure}[htbp]
    \centering
        \includegraphics[scale=.24]{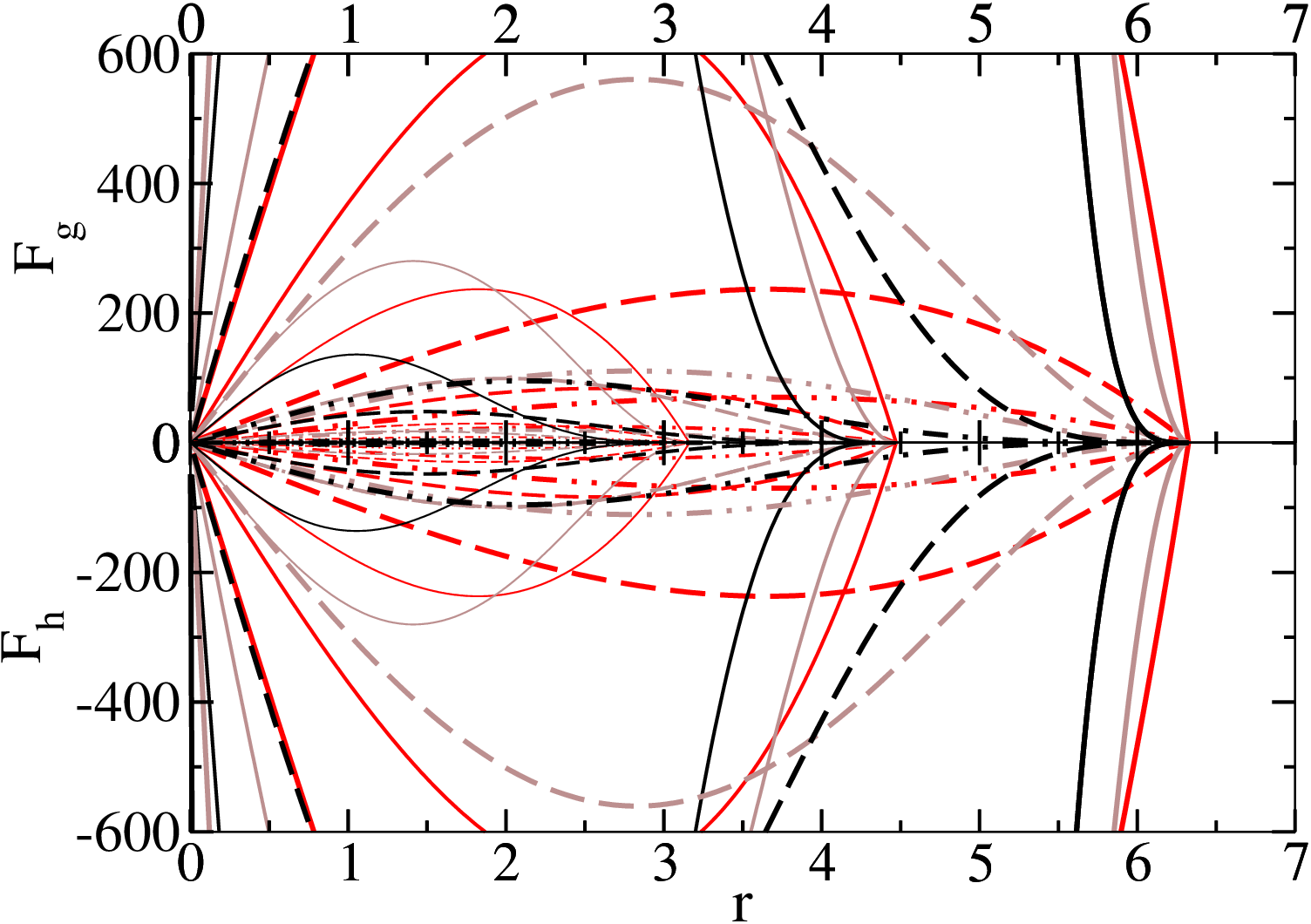}
       \caption{  Variation of the forces    in the
  interior region. Description of curves is the same as in FIG.~\ref{fig:1}.
}
    \label{fig:2}
\end{figure}

The compactness of the fluid sphere, $u(r)$, is thus defined as be
found as
\begin{equation}
\label{eq30} u(r)=  M(r)/r.
\end{equation}
This is an increasing function of the radial parameter ( see figure
3). Correspondingly,
 the surface redshift ($Z_s$)  is given by
\begin{equation}
\label{eq34} Z_s(r)= ( 1-2 u(r))^{-\frac{1}{2}} - 1.
\end{equation}

FIG.~4 provides variation of $Z_s$ against r for different values of
the parameters.

In (2+1) dimensional spacetime, the vacuum solution does not exist
without cosmological constant. Thus it is not possible
to match our interior solution with the BTZ black hole as it is the
vacuum solution with non zero $\Lambda$. However, if one takes B as
large as possible, then the  solution is valid for the infinite large
fluid sphere. This means that  we don't have the vacuum region left.
\\
\begin{figure}[htbp]
    \centering
        \includegraphics[scale=.24]{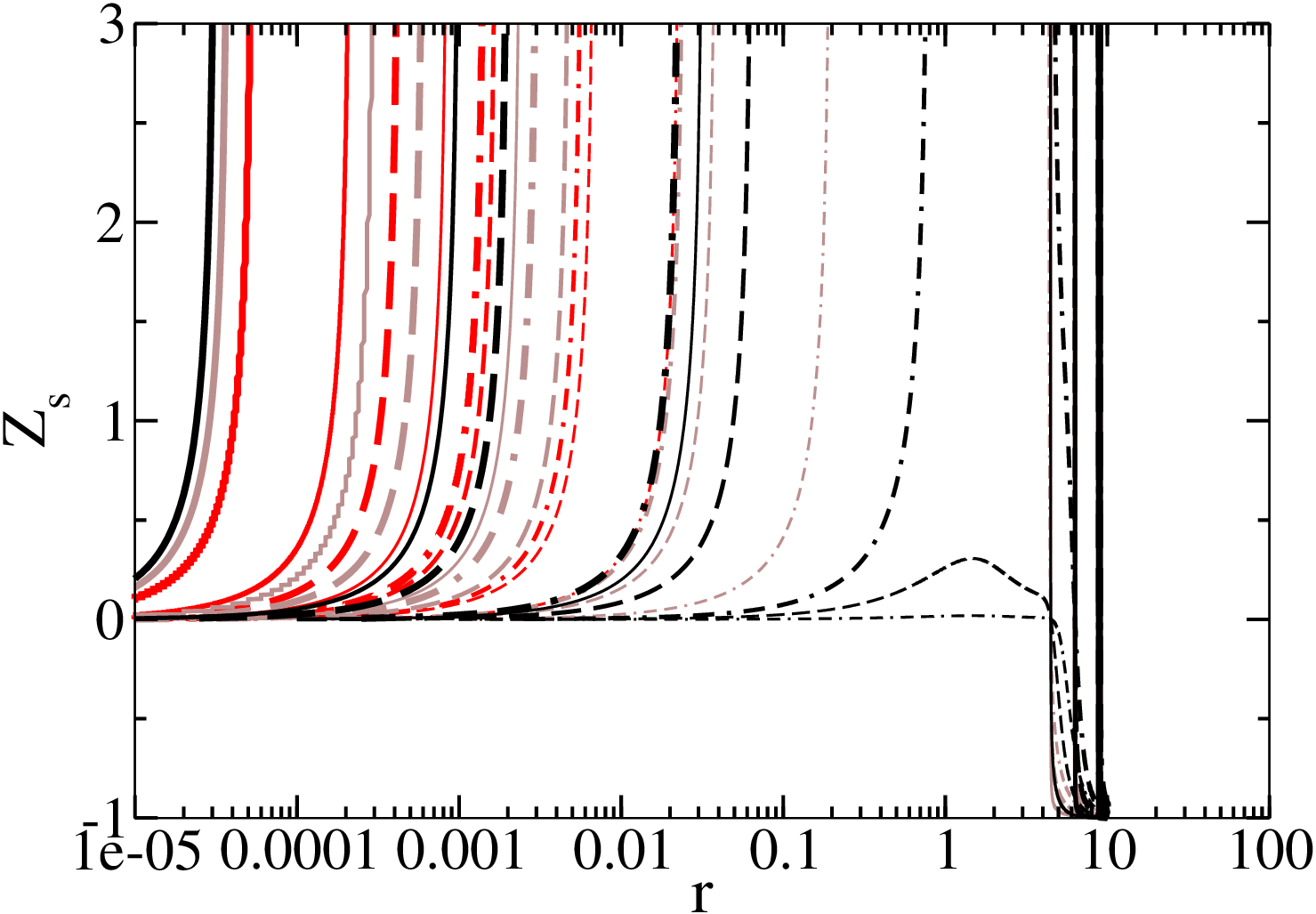}
       \caption{  Variation of redshift in the
  interior region.  Description of curves is same as in FIG.~\ref{fig:1}.}
    \label{fig:5}
\end{figure}

\subsubsection{$~m=1~$}
For stiff  fluid model, $p=\rho$, and with $\Lambda= 0$, the field
equations (3)-(6) yield  following solutions
 \begin{eqnarray}
\nu &=& e^{-D}(r^{2}/2)+E,\\
\mu &=& -D + e^{-D}(r^{2}/2) + E, \\
\rho&=& F e^{-2E-r^{2}e^{-D}}.
\end{eqnarray}
Here $D$, $E$ and $F$ are integration constants.

For the consistency of the solutions, the constants should follow
the following constraint equation.
\begin{equation}\label{r40} D= \ln(2 \pi F)~~. \end{equation}
This ensures that $F>0$. \\The solutions are regular at the center
and are valid for infinite large sphere. The central density is
$\rho_c= F e^{-2E}$.
\\
\begin{figure}[htbp]
    \centering
        \includegraphics[scale=.24]{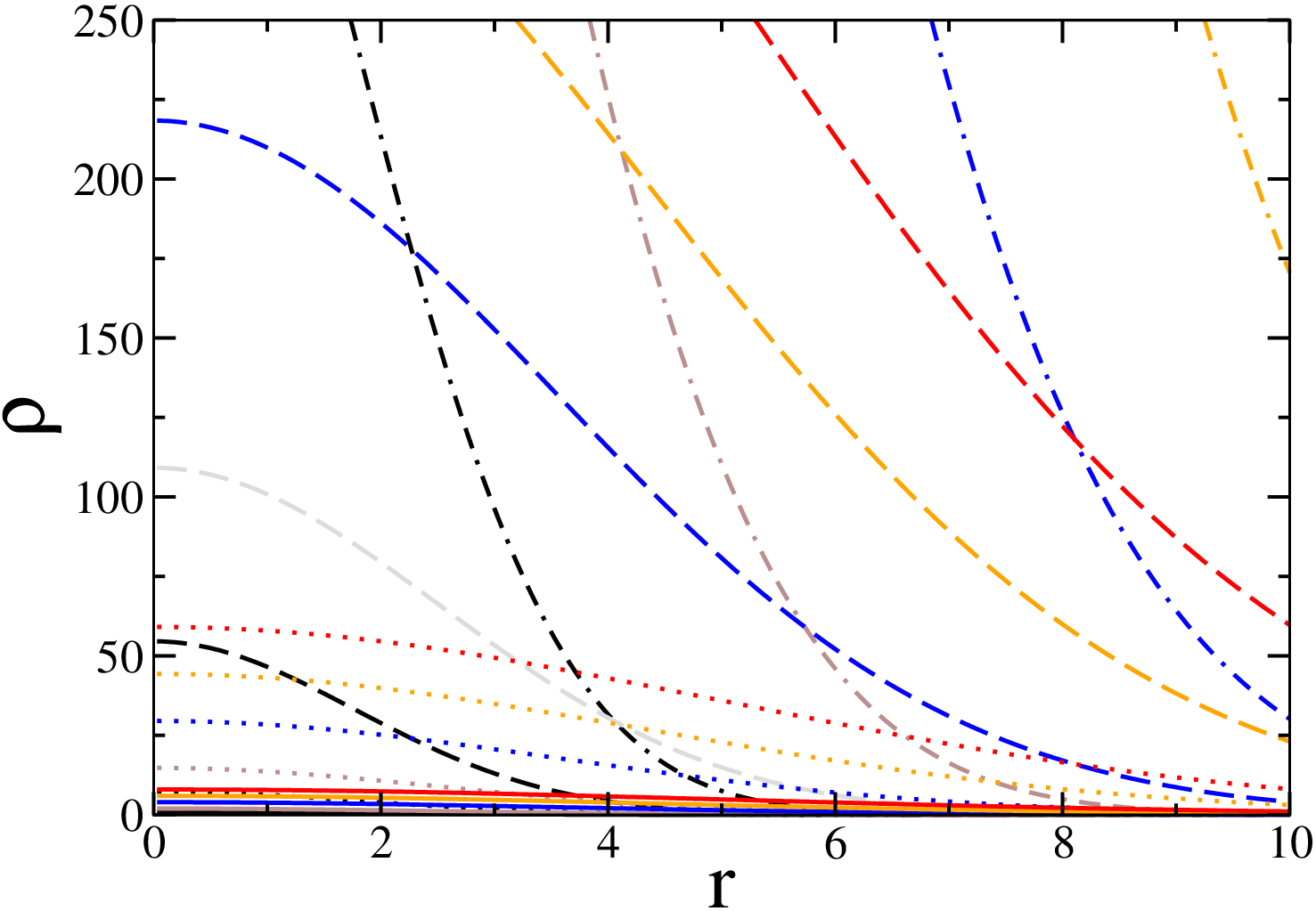}
       \caption{  Variation of the energy-density($\rho$)    in the
  interior region for m=1. Description of curves are as follows:
   $ F=1,2,4,6$ and $8$  correspond to black, brown, blue, orange and red colors, respectively.
$E=0, -1, -2$ and $-3$  correspond to solid, dotted, dashed and
dot-dashed lines, respectively. For $ E\geq0$ values are too small.
}
    \label{fig:5}
\end{figure}
\\
\\
\\

From  Eq.~(32), we find at $r=0$,
\begin{equation}
\frac{dp}{dr}=0,~~ \frac{d\rho}{dr} = 0~~{\rm and}~~
\frac{d^{2}\rho}{dr^{2}}=\frac{1}{\pi}\left[-e^{-2E}\right] <0
\end{equation}
Thus central density is maximum.
\\

The mass, $M(r)$,  within a  radial distance $r$ is given by

\begin{equation}
M(r)=e^{D}\left[e^{D-2E}-e^{D-2E-r^{2}e^{-D}}\right]/2~~.
\end{equation}
The compactness of the fluid sphere  is thus, $u(r)= M(r)/r$. Having
$u$, the $Z_s$ is determined using Eq.~(\ref{eq34}). The important
physical characteristics such as density, compactness  and redshift
are shown in FIGs. 5-7.\\
\\
\begin{figure}[htbp]
    \centering
        \includegraphics[scale=.24]{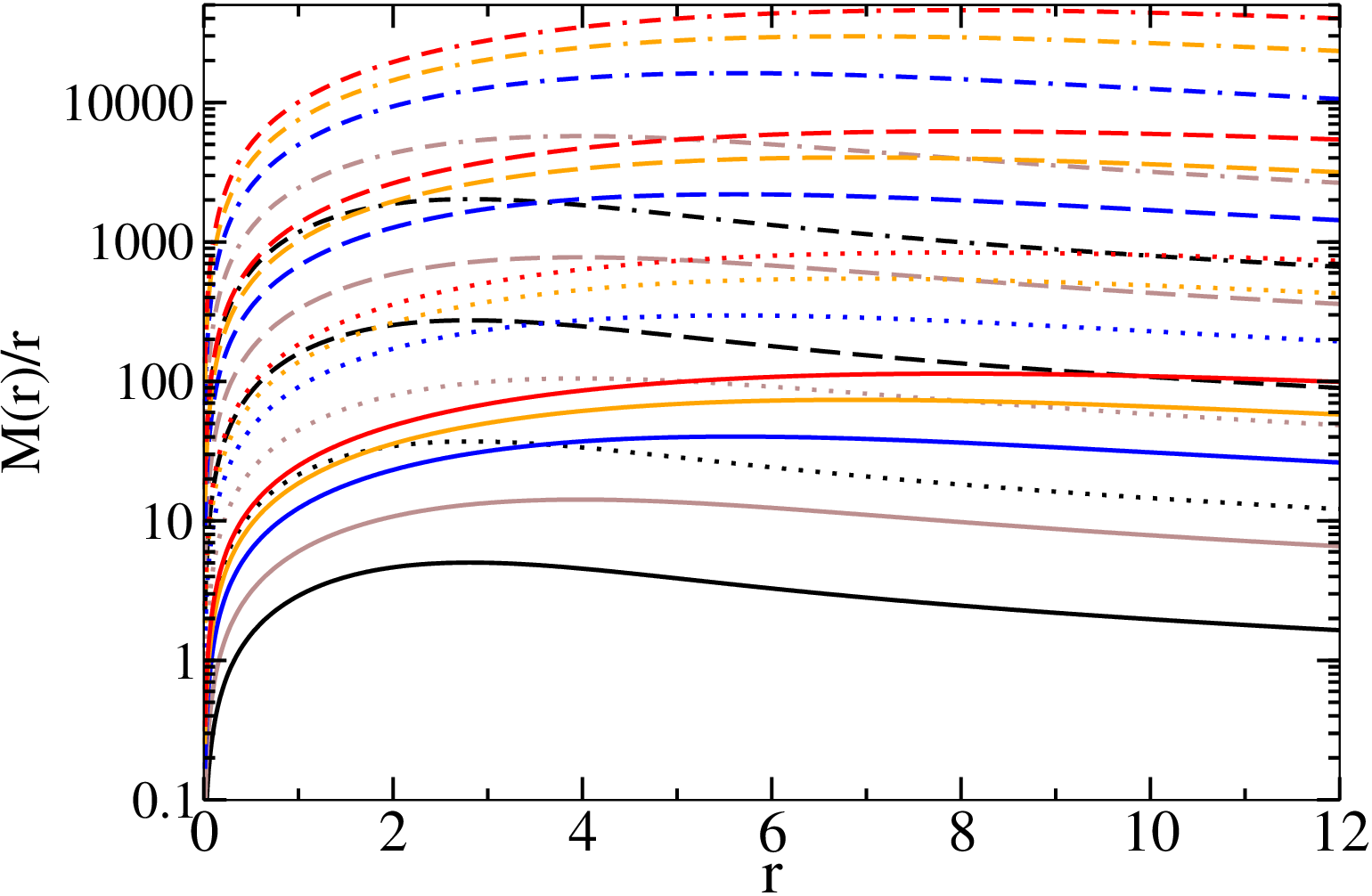}
       \caption{    Variation of compactness( $u=\frac{M(r)}{r}$  )  in the
  interior region. Description of curves is the same as in FIG.~5. }
    \label{fig:2}
\end{figure}
\\

 The TOV equation  yields
\begin{equation}
 F_g+ F_h =0,
\end{equation}
where
 \begin{equation}
F_g=-F_h=(r/\pi)e^{-2E-r^{2}e^{-D}}.\end{equation}
\\
\\
\begin{figure}[htbp]
    \centering
        \includegraphics[scale=.24]{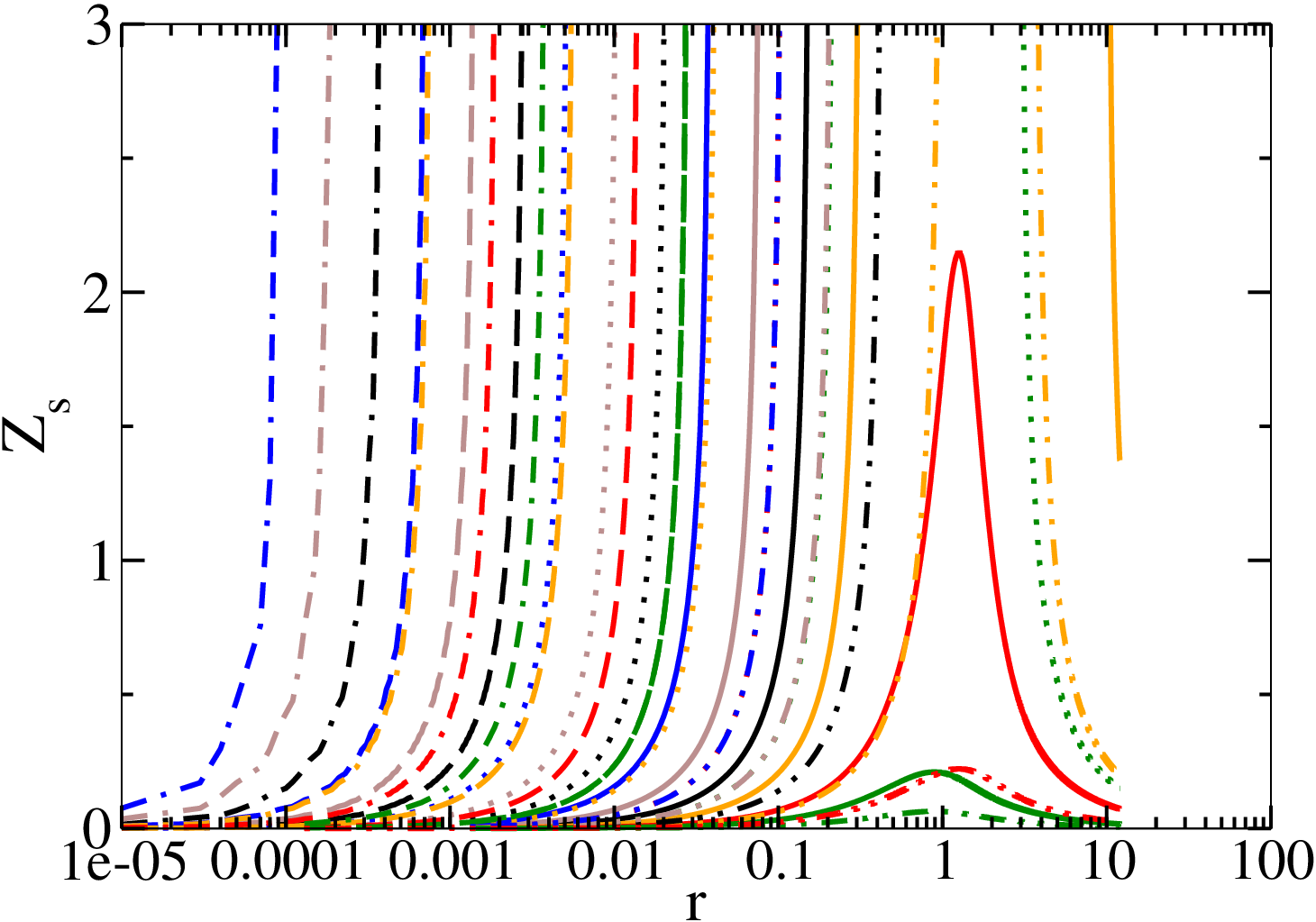}
       \caption{  Variation of redshift in the
  interior region.  Description of curves is the same as in FIG.~4. }
    \label{fig:9}
\end{figure}

 The profiles of
$F_g$ and  $F_h$  for the specific values of parameters are shown in
FIG.~8, which provides information about the static equilibrium
due to gravitational and hydrostatic forces combined. As before, we
can not match our interior solution with BTZ exterior vacuum
solution.
\\
\\
\begin{figure}[htbp]
    \centering
        \includegraphics[scale=.24]{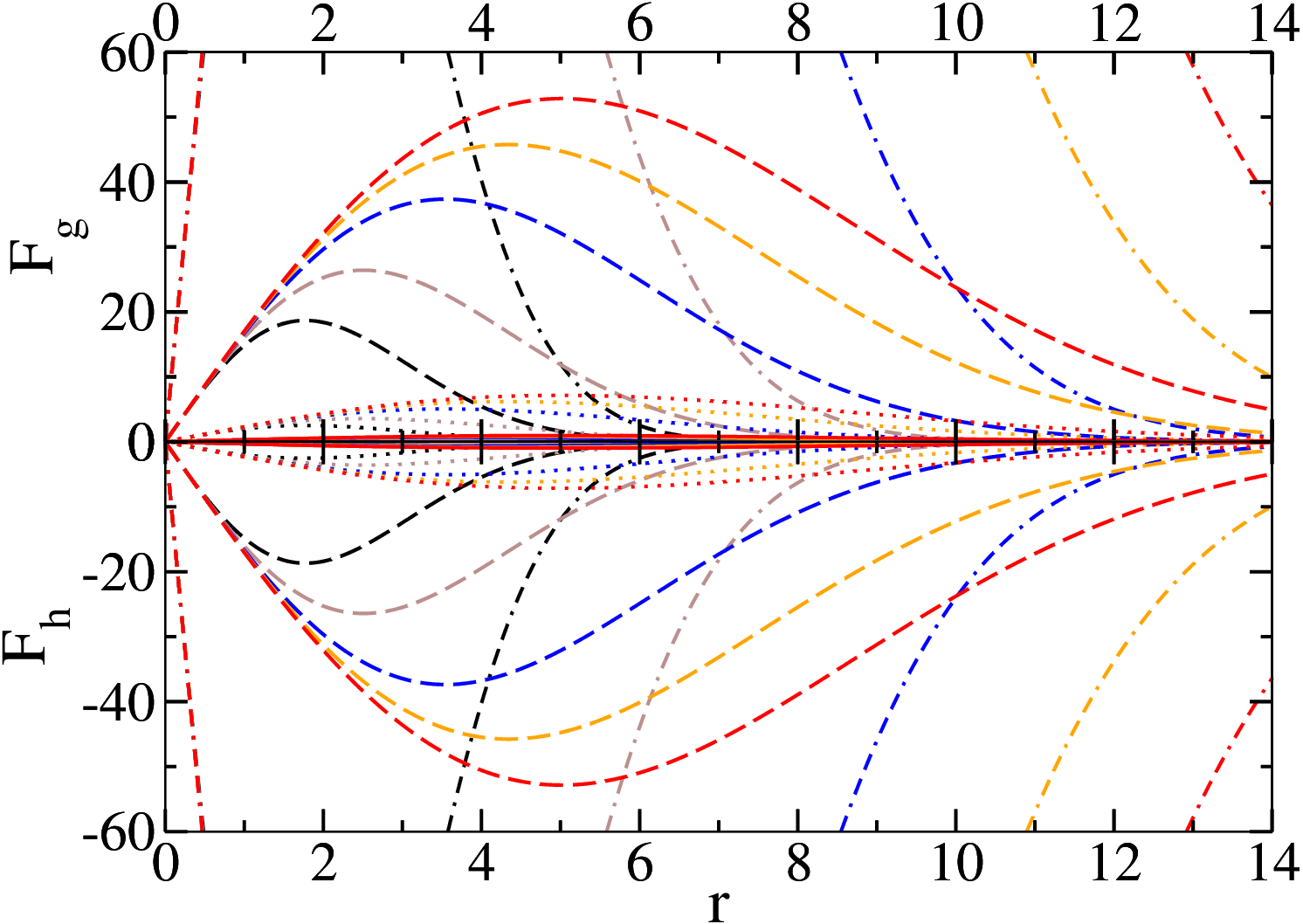}
       \caption{   Variation of the forces    in the
  interior region for $m$=1. Description of curves is the same as in
  FIG.~4.
}
    \label{fig:10}
\end{figure}

 \subsubsection{$~m=-1~$}
The equation of state of the kind, $p =-\rho$   is related to the
$\Lambda-dark \:\: energy$, an agent responsible for the second
phase of the inflation of Hot Big Bang theory. Using  the equation
of state of the kind, $p =-\rho$, and with $\Lambda= 0$, the field
equations (3)-(6)  yield following solutions

\begin{eqnarray}
\rho&=& -p = J,\\
\nu &=& \left[H+ \ln (r^{2}+K)\right]/2,\\
\mu &=& \left[H- \ln (r^{2}+K)\right]/2.
\end{eqnarray}

Here $J$, $K$  and $H$ are integration constants.
Solutions hold good for the following constraint equation
\begin{equation}2 \pi J + e^{-H}=0~~.
\end{equation}
These  are regular at the center if $K$ is positive and the solution
is valid for the infinite large fluid sphere. However, for $K<0$,
solution is valid for $r>\sqrt{-K}$ up to infinite large radius.

\subsubsection {$~m=0~$}
For the dust case i.e. when $p=0$ and $\rho \neq 0$, the field
equations (3)-(6) reduce to
\begin{equation}\nu = \nu_0
\end{equation}
and
\begin{equation}e^{-2\mu} = \mu_0 - \int 4 \pi r \rho dr.
\end{equation}
Here, $\nu_0 ~and~ \mu_0$ are integration constants.

Unless   specifying the energy density, one can not get
exact analytical solution of the field equations. Thus dust model in
(2+1) dimensional space time is   possible for known energy density.
\subsection{With cosmological constant $(\Lambda\neq 0)$}
 \subsubsection{$~m=-1~$}
As before for  the equation of state of the kind $p =-\rho$ with non
zero $\Lambda$, the field equations (3)-(6) yield $\rho = c_4$. The
metric coefficients  may be obtained as
  \begin{equation}\nu = \ln(r^{2}+B_5)/2+D_5 \end{equation}
and
 \begin{equation}\mu = -\ln (r^{2}+B_5)/2 -D_5 +A_4. \end{equation}
 Here, $c_4$, $A_4$, $B_5$ and $D_5$ are integration constants.

These solutions are consistent if
 \begin{equation}2 \pi c_4 +\Lambda + e^{2D_5-2A_4}=0~~.\end{equation}
These solutions  are regular at the center if $B_5$ is positive and
the solution is valid for the infinite large fluid sphere. However,
for $ B_5 <0$, the   solution is valid for $r>\sqrt{-B_5}$ up to
infinite large radius. The nature of the solutions of the metric
potentials is independent of the sign of  $\Lambda$. However, sign
of $\Lambda$
plays a crucial role to get positive energy density.
For positivity of energy density, one should take negative $\Lambda$.

Note that without any loss of generality, we can take $D_5 =0$ as it
can be absorbed by re-scaling the coordinates.
We match the interior solution to the exterior BTZ black hole metric
\[ds^{2}=-(-M_0-\Lambda r^{2})dt^{2}+(-M_0-\Lambda r^{2})^{-1}dr^{2}+r^{2}d\theta^{2},\]
at the boundary $r=R$, which yield
  \begin{equation}\left(-M_0-\Lambda R^{2}\right)=R^{2}+B_5 ,\end{equation}
 \begin{equation}\left(-M_0-\Lambda
R^{2}\right)^{-1}=e^{2A_4}(R^{2}+B_5)^{-1}.\end{equation} Solving
these two  equations, we get
 \begin{equation}B_5=-M_0-(\Lambda +1 )R^{2}~~and ~~A_4=0~~.\end{equation}
The consistency relation assumes the form
 \begin{equation}2 \pi c_4 +\Lambda + 1=0~~.\end{equation}

\subsubsection{ $0<m\leq 1$}
From the field equations (3)-(6) after some manipulation, we arrive
at
\begin{equation}\label{51}2 \pi (1+m)A_1
e^{\frac{m-1}{m}\nu}\nu'^{3}=2r\nu'^{2}+r\nu''-\nu'~~.\end{equation}

One can observe   that $\nu'=0$ will be a particular solution of
this equation. This yields $\nu=constant$. Equation (4) implies $p =
\Lambda/2 \pi$. Finally, we  get the following solution for  $\mu$
as
\begin{equation}
\mu=-\ln\left[A_5-Nr^2\right]/2~~.
\end{equation}
Here, $N=\Lambda (1+m)/m$ and  $A_5$ is integration constant.

For positivity of energy density, one should take positive
$\Lambda$. The solutions  are regular at the center if $A_5>0$ and
valid up to $r< \sqrt{A_5/N}$.
In this case, we can not match the interior solution to the exterior
BTZ black hole metric which is vacuum solution with negative
$\Lambda$.

\subsubsection{$~m=0~$}
For the dust case i.e. when $p=0$ and $\rho \neq 0$,  one can not
obtain the exact analytical solution of the field equations. Thus
dust model in $(2+1)$ dimensional spacetime with non zero $\Lambda$
is not possible.

 \section{ Conclusion }\label{conclusion}
In this paper we have obtained a new class of exact  interior
solution of Einstein field equation in $(2+1)$ dimensional space
time   assuming the equation of state  $p=m \rho $ ( where $ m $ is
the equation of state parameter).  The interior solutions obtained without
cosmological, $\Lambda$,  are physically acceptable for the
following reasons:\\
  (i) the solutions are  regular at the origin,\\
  (ii)  both the
pressure (p) and energy density ($\rho$) are  positive definite at
the origin,\\
 (iii)  the pressure reduce to zero at some finite
boundary radius $r_b > 0$,\\
 (iv) both the pressure and energy density
are monotonically decreasing to the boundary,\\
 (v) the subluminal
sound speed ($v^2_s =\frac{ dp }{d \rho} = m \leq 1$)\\
(vi) and Ricci scalar is non zero i.e. spacetime is non flat.

 It is to be noted that at very high densities the adiabatic sound
speed may not equal the actual propagation speed of the signal. By
studying TOV equation, we have shown that  equilibrium stage of the
interior region without  $\Lambda$ can be achieved due to the
combined effect of gravitational and hydrostatic forces. We know BTZ
exterior vacuum solution in (2+1) dimension is valid only for non
zero $\Lambda$. Therefore, it is not possible to match our interior
solution ( without $\Lambda$ ) with BTZ spacetime at some boundary.
We emphasis the following fact that any interior solution in four
dimensional space  made with a perfect fluid must be glued with an
exterior vacuum solution only at a regular surface p = 0 ( this is
consequence of the well-known Israel matching conditions for the
related problem ).  For barotropic equation of state the
configurations present p = 0 surfaces at the same location where
$\rho = 0$. For the solutions (15)-(17) with $0 < m < 1$  this
occurs at $r = \sqrt{2B}$. However, the metric coefficients are
singular at the same locus, in fact, there is a curvature
singularity at $r = \sqrt{2B}$.  Hence, this is not a regular region
where spacetime can be continuously glued with other spacetime.
Hence,  one should take B as large as possible so that  the
solution is valid for the infinite large fluid sphere and  we don't
have the vacuum region left. For m=1 case, there does not exist any
radius for which p or $\rho=0$, hence, Israel matching condition
does not occur.

While finding  interior solution with non zero $\Lambda$, we note that
density and pressure remain constant. Interestingly, we observe that
it is not possible to get dust model in $(2+1)$ dimensional
spacetime with non zero $\Lambda$. Investigation on full collapsing
model of a (2 + 1) dimensional configuration will be a future
project.

\section{Acknowledgements}
FR , AAU and RB would like to thank the Inter-University Centre for
Astronomy and Astrophysics (IUCAA), Pune, India, for research
facility.  FR is also grateful to UGC, Govt. of  India, for
financial support under its Research Award Scheme. PB is  thankful
to CSIR, Govt. of  India for providing JRF. RB thanks CSIR for
awarding Research Associate fellowship. We are thankful to Dr. R.
Sharma for valuable discussion.


 \end{document}